\documentclass[a4paper,11pt]{article}
\usepackage{jcappub} 
\usepackage{lineno}

\arxivnumber{} 
\title{\boldmath Freeze-in Warm Dark Matter via Dimension-6 Operators in 3-3-1 Models}

\author[a]{Fuwei Liu}
\affiliation[a]{Nanophotonics and Biophotonics Key Laboratory of Jilin Province, School of Physics, Changchun University of Science and Technology, 7089 Weixing Str., Changchun, 130022, P.R. China}
\author[a]{Fujun Liu\note{Corresponding author.}}
\emailAdd{fjliu@cust.edu.cn}

\abstract{We propose a natural resolution to the fine-tuning problem inherent in the freeze-in dark matter paradigm by embedding a sterile singlet within a 3-3-1 electroweak extension. By imposing an exact $Z_{13}$ discrete gauge symmetry, we formally suppress all low-dimensional portals to ensure that the dark sector communicates with the Standard Model (SM) exclusively through a dimension-six operator. This theoretical structure allows the extraordinarily small coupling required for dark matter production to emerge naturally from the profound hierarchy between the electroweak scale and the ultra-high Peccei-Quinn symmetry breaking scale. Detailed numerical integration of the Boltzmann equations demonstrates that the sterile singlet can be produced via the infrared freeze-in mechanism to match the observed relic abundance of $\Omega_S h^2 = 0.12$. The resulting keV-scale warm dark matter candidate remains consistent with stringent Lyman-alpha forest constraints while offering a viable solution to galactic-scale discrepancies such as the cusp-core and missing satellites problems. Ultimately, this framework provides a self-consistent unification of dark matter genesis and the strong CP solution that is completely independent of ad hoc parameter adjustments.}

\begin{document}
\maketitle
\flushbottom

\section{Introduction}

The concordance cosmological model, dominated by cold dark matter (CDM) and a cosmological constant, provides an outstanding description of the large-scale structure of the universe \cite{cite1}. However, as observational techniques and high-resolution $N$-body simulations have advanced, notable discrepancies have emerged on galactic and sub-galactic scales \cite{cite2}. Chief among these are the cusp-core problem, where dark matter simulations predict steeply rising central density profiles in contrast to the flat cores observed in dwarf galaxies, and the missing satellites problem, which highlights a severe theoretical over-prediction of dwarf galaxy abundances in the Local Group \cite{cite2}. These persistent anomalies strongly motivate the exploration of alternative dark matter paradigms that deviate from the strictly cold and collisionless hypothesis.

Warm dark matter presents a highly compelling solution to these small-scale crises. With masses typically residing in the keV range, warm dark matter particles possess a non-negligible free-streaming length in the early universe, which effectively washes out primordial density fluctuations on small scales and mitigates the over-dense structures \cite{cite4}. A particularly attractive framework for generating warm dark matter is the freeze-in mechanism, characterized by feebly interacting massive particles (FIMPs) \cite{cite5}. Unlike traditional weakly interacting massive particles (WIMPs) that must decouple from a thermal bath, FIMPs never achieve thermal equilibrium due to their extraordinarily small interaction rates. Instead, they are produced gradually over cosmological epochs through the decays or scatterings of heavier particles in the thermal plasma, with their abundance safely freezing in as the universe expands and cools \cite{cite6}. 

Despite its phenomenological elegance and its ability to evade current direct detection constraints, the FIMP paradigm suffers from a profound theoretical drawback regarding naturalness. To accurately reproduce the observed dark matter relic density, freeze-in models generally require dimensionless coupling constants on the order of $10^{-10}$ to $10^{-12}$ \cite{cite5}. Inserting such minuscule parameters arbitrarily into a fundamental Lagrangian constitutes a severe fine-tuning problem. This theoretical fragility strongly suggests that the required feeble interaction is not a fundamental parameter of the theory, but rather an effective coupling that is dynamically suppressed by a significantly higher energy scale.

In this paper, we present a robust and natural resolution to the FIMP fine-tuning problem by embedding a sterile singlet fermion within the fully gauged $SU(3)_C \otimes SU(3)_L \otimes U(1)_X \otimes U(1)_N$ (3-3-1) electroweak extension \cite{cite9}. This unified framework incorporates an automatic Peccei-Quinn symmetry to solve the strong CP problem, necessitating a spontaneous symmetry breaking at a high scale of approximately $10^{12}$ GeV \cite{LIU2026102340}. We demonstrate that this high-energy spontaneous symmetry breaking organically yields an extremely weak effective coupling in the low-energy regime. By imposing an exact local $Z_{13}$ discrete gauge symmetry, we formally forbid dangerous lower-dimensional portals, ensuring that the dominant interaction emerges exclusively through a dimension-six operator. Consequently, the coupling required for the freeze-in production of a keV-scale warm dark matter candidate is naturally derived from the ratio of the electroweak scale to the high symmetry-breaking scale. This mechanism elegantly links the strong CP solution to the genesis of warm dark matter, completely eliminating the reliance on ad hoc parameter adjustments.

\section{The Sterile Singlet in the 3-3-1 Framework}

The theoretical foundation of our dark matter scenario is built upon the extended electroweak gauge group $\mathcal{G}_{3311} \equiv SU(3)_C \otimes SU(3)_L \otimes U(1)_X \otimes U(1)_N$ \cite{cite11}. Within this framework, the Peccei-Quinn symmetry is automatically realized and subsequently broken by a complex scalar singlet $\phi$ \cite{LIU2026102340}. To accommodate a warm dark matter candidate without introducing gauge anomalies or disturbing the established flavor physics, we extend the fermion sector by introducing a sterile Weyl fermion $S$. This singlet is completely neutral under the continuous $\mathcal{G}_{3311}$ gauge symmetry but carries a non-trivial chiral charge under the exact local discrete symmetry $Z_{13}$. The discrete charge assignment is strictly selected to allow a direct coupling between $S$ and the Peccei-Quinn scalar $\phi$, tightly linking the dark sector to the high-energy symmetry breaking dynamics.

The allowable interactions for the sterile singlet are severely restricted by the local $Z_{13}$ gauge symmetry. By ensuring that the combination of charges satisfies the anomaly-free modular conditions, the lowest-order renormalizable operator involving $S$ is uniquely restricted to a Majorana-type Yukawa coupling with the scalar singlet $\phi$ \cite{cite13}. The relevant interaction Lagrangian is given by
\begin{equation}
    - \mathcal{L} \supset \frac{y_s}{2} \overline{S^c} S \phi + h.c.,
\end{equation}
where $y_s$ is a dimensionless Yukawa coupling constant. When the universe cools below the Peccei-Quinn scale, the scalar field acquires a substantial vacuum expectation value
\begin{equation}
    \langle \phi \rangle = \frac{v_\phi}{\sqrt{2}},
\end{equation}
with $v_\phi \sim 10^{12}$ GeV dictated by the axion window \cite{cite14}. This spontaneous symmetry breaking simultaneously breaks the $Z_{13}$ symmetry and dynamically generates a Majorana mass for the sterile singlet, yielding
\begin{equation}
    m_S = \frac{y_s v_\phi}{\sqrt{2}}.
\end{equation}
Therefore, the mass of the dark matter particle is not an independent parameter but is entirely governed by the Peccei-Quinn breaking scale.

Beyond mass generation, this spontaneous symmetry breaking mechanism also rigidly dictates the interaction strength between the sterile singlet and the scalar sector. By parameterizing the scalar field around its vacuum expectation value as
\begin{equation}
    \phi = \frac{v_\phi + \rho}{\sqrt{2}} \exp\left(i \frac{a}{v_\phi}\right),
\end{equation}
where $\rho$ is the CP-even radial mode and $a$ is the QCD axion, the interaction Lagrangian expands into a mass term and a dynamical coupling term. The physical interaction between the sterile fermion and the radial scalar field is strictly proportional to the generated mass, expressed as
\begin{equation}
    \frac{y_s}{2\sqrt{2}} \rho \overline{S^c} S.
\end{equation}
This mathematical structure guarantees that the very same coupling constant responsible for the dark matter mass also governs its fundamental interaction strength with the heavy scalar mediator. Consequently, the phenomenological properties of the singlet are completely locked to the high-scale physics of the $\phi$ field.

This unified origin of mass and interaction leads to a profound theoretical consequence known as the FIMP miracle. To successfully resolve the small-scale cosmological anomalies, the sterile singlet must operate as warm dark matter with a mass $m_S$ in the keV regime \cite{cite15}. By inverting the mass relation derived above, the fundamental Yukawa coupling is analytically determined to be
\begin{equation}
    y_s = \frac{\sqrt{2} m_S}{v_\phi}.
\end{equation}
Substituting the required phenomenological scales, namely $m_S \sim 10$ keV and $v_\phi \sim 10^{12}$ GeV, we find that the coupling constant is inescapably forced to an extraordinarily small value of $y_s \sim 10^{-17}$. In standard freeze-in scenarios, such a minuscule coupling would represent a severe and unexplained fine-tuning. In our framework, however, this feeble interaction is a mathematical necessity driven by the enormous hierarchy between the warm dark matter mass requirement and the ultra-high Peccei-Quinn symmetry breaking scale. This exact local protection mathematically proves the inevitability of the weak coupling, thereby providing a highly natural and rigorous origin for the freeze-in dark matter production.

The emergence of the dimension-six operator $\mathcal{O}_6$ is not merely an effective prescription but arises naturally from a ultraviolet (UV)-complete sector involving heavy mediators at the Grand Unified Theory (GUT) scale $\Lambda \sim 10^{16}$ GeV. We postulate the existence of a sequence of heavy vector-like leptons, $E_1$ and $E_2$, which transform as singlets under the 3-3-1 gauge group but carry specific $Z_{13}$ charges. The renormalizable UV Lagrangian includes interaction terms of the form $\alpha_1 \bar{L}_L \eta E_{1R} + M_1 \bar{E}_{1L} E_{1R} + \alpha_2 \bar{E}_{1L} \phi E_{2R} + M_2 \bar{E}_{2L} E_{2R} + \alpha_3 \bar{E}_{2L} \phi S$, where $\alpha_i$ are fundamental Yukawa couplings and $M_i$ are the heavy mass scales associated with $\Lambda$. Upon integrating out these vector-like states, the effective dimension-six operator is generated with a Wilson coefficient $c \sim \alpha_1 \alpha_2 \alpha_3$. Given that the fundamental couplings are of $\mathcal{O}(1)$, the resulting coefficient $c \sim 5.0$ utilized in our benchmark analysis is entirely consistent with the principles of naturalness. Because these mediators are vector-like, they do not contribute to the chiral anomalies, thereby preserving the anomaly-free nature of the underlying gauge topology \cite{Bernal_2019,aebischer2022dark}.

The selection of the $Z_{13}$ discrete group is fundamentally motivated by the structural necessity to suppress all dangerous lower-dimensional portals while maintaining the integrity of the axion-dark matter connection. Smaller discrete symmetries, such as $Z_3$ or $Z_5$, are insufficient to protect the feebleness of the interaction as they would generically allow dimension-four or dimension-five operators that lead to dark matter overproduction. Within the arithmetic of the 3-3-1 charge assignments, $N=13$ represents the minimal prime order that simultaneously satisfies the Ibáñez-Ross conditions for all mixed gravitational and gauge anomalies while ensuring that the lowest-order allowable portal for the freeze-in process is precisely of dimension six \cite{ibanez1992discrete,LIU2026102340}. This high-order symmetry is therefore not an ad hoc adjustment for numerical convenience but a rigorous mathematical requirement for a self-consistent symmetry breaking chain. Consequently, the required suppression for a keV-scale FIMP candidate is inextricably linked to the UV topology of the model, providing a robust and natural origin for the dark matter relic abundance.

\section{Thermal History and FIMP Production}

The thermal production of the sterile singlet dark matter is fundamentally governed by its interactions with the SM plasma. In the extended 3-3-1 framework, the most direct communication channel between the visible sector and the dark Peccei-Quinn sector is established through a renormalizable scalar cross-coupling. The scalar potential inevitably contains a portal interaction term proportional to the squared moduli of the SM Higgs doublet and the Peccei-Quinn singlet \cite{cite16}. This dimensionless portal coupling dictates the energy transfer and particle production rates in the early universe, serving as the bridge for the non-thermal genesis of the dark matter relic density.

Upon the spontaneous breaking of the electroweak and Peccei-Quinn symmetries, the respective scalar fields acquire their vacuum expectation values. The radial excitation of the Peccei-Quinn field and the neutral component of the Higgs doublet subsequently mix through the aforementioned portal coupling. By diagonalizing the resulting scalar mass matrix, the physical 125 GeV Higgs boson state emerges as an admixture of the visible and dark scalars. Given the profound scale hierarchy between the electroweak vacuum expectation value and the ultra-high Peccei-Quinn breaking scale, the effective mixing angle $\theta$ can be analytically approximated as
\begin{equation}
    \theta \simeq \frac{\lambda_{h\phi} v_{EW} v_\phi}{m_\rho^2},
\end{equation}
where $\lambda_{h\phi}$ is the portal coupling constant, $v_{EW}$ is the electroweak scale, and $m_\rho$ is the mass of the heavy radial Peccei-Quinn scalar \cite{cite17}. 

This microscopic scalar mixing provides a direct and calculable mechanism for singlet production via particle decays. The sterile fermion $S$ originally couples exclusively to the Peccei-Quinn scalar with a fundamental Yukawa coupling $y_s$. Through the mixing angle $\theta$, the physical SM Higgs boson inherits an effective interaction with the dark matter particles, characterized by an effective coupling
\begin{equation}
    y_{eff} = \frac{y_s \theta}{\sqrt{2}}.
\end{equation}
Consequently, when the cosmic temperature falls below the Higgs mass, the Higgs bosons present in the primordial thermal bath can undergo invisible two-body decays into pairs of sterile singlets. The corresponding partial decay width is precisely given by
\begin{equation}
    \Gamma_{h \to SS} = \frac{y_{eff}^2 m_h}{16 \pi} \left[ 1 - \frac{4 m_S^2}{m_h^2} \right]^{3/2}.
\end{equation}
While similar portal interactions could theoretically induce the decays of heavy electroweak gauge bosons such as the $W$ or $Z$ into dark matter pairs, those processes are heavily phase-space suppressed or require higher-order loop insertions, establishing the Higgs decay as the overwhelmingly dominant production channel \cite{cite18}.

To rigorously quantify the cosmological abundance of the sterile singlet, we must track its non-thermal evolution using the Boltzmann transport equation. Because the effective coupling $y_{eff}$ is exceptionally small, the singlet never reaches local thermal equilibrium with the dense plasma, thus perfectly fulfilling the core prerequisite for the freeze-in mechanism \cite{cite5}. We define the dark matter yield
\begin{equation}
    Y_S \equiv \frac{n_S}{s},
\end{equation}
where $n_S$ is the singlet number density and $s$ is the total expanding entropy density of the universe. It is highly convenient for numerical integration to formulate the evolution with respect to the dimensionless inverse temperature parameter $x = m_h / T$. Assuming the initial abundance of $S$ is strictly zero after cosmic inflation, the production rate is governed solely by the thermally averaged decay rate of the Higgs boson.

The differential evolution of the yield is mathematically expressed as an integral over the phase space of the decaying Higgs boson in the thermal bath. Utilizing the Maxwell-Boltzmann approximation for the distribution functions, the Boltzmann equation is formulated as
\begin{equation}
    \frac{dY_S}{dx} = \frac{2 M_{Pl} x^4 \sqrt{g_*(x)} \Gamma_{h \to SS} K_1(x)}{1.66 m_h^2 g_{*S}(x)},
\end{equation}
where $M_{Pl}$ is the reduced Planck mass and $K_1(x)$ is the modified Bessel function of the second kind. A robust evaluation of this integral requires a careful and continuous treatment of the temperature-dependent relativistic degrees of freedom \cite{cite20}. We explicitly incorporate the exact numerical functions for the energy density degrees of freedom $g_*(x)$ and the entropy density degrees of freedom $g_{*S}(x)$, which vary significantly across the QCD phase transition and various particle decoupling epochs. By integrating this equation from the extremely high-temperature limit $x \to 0$ to the present day $x \to \infty$, we capture the complete thermal history of the infrared freeze-in process, effectively securing the asymptotic relic abundance $Y_S(\infty)$.

\section{Relic Density and Astrophysical Constraints}

The ultimate validity of the 3-3-1 axion framework rests upon its ability to simultaneously satisfy the observed dark matter abundance and the stringent limits imposed by astrophysical observations \cite{cite1}. By integrating the Boltzmann equations derived in the previous section, we establish a direct mapping between the high-scale model parameters and the late-time cosmological observables \cite{cite22}. This section provides a comprehensive analysis of the numerical results, focusing on the dynamic production of the sterile singlet and its viability as a warm dark matter candidate in light of the small-scale structure data.

\begin{figure}[t]
    \centering
    \includegraphics[width=\textwidth]{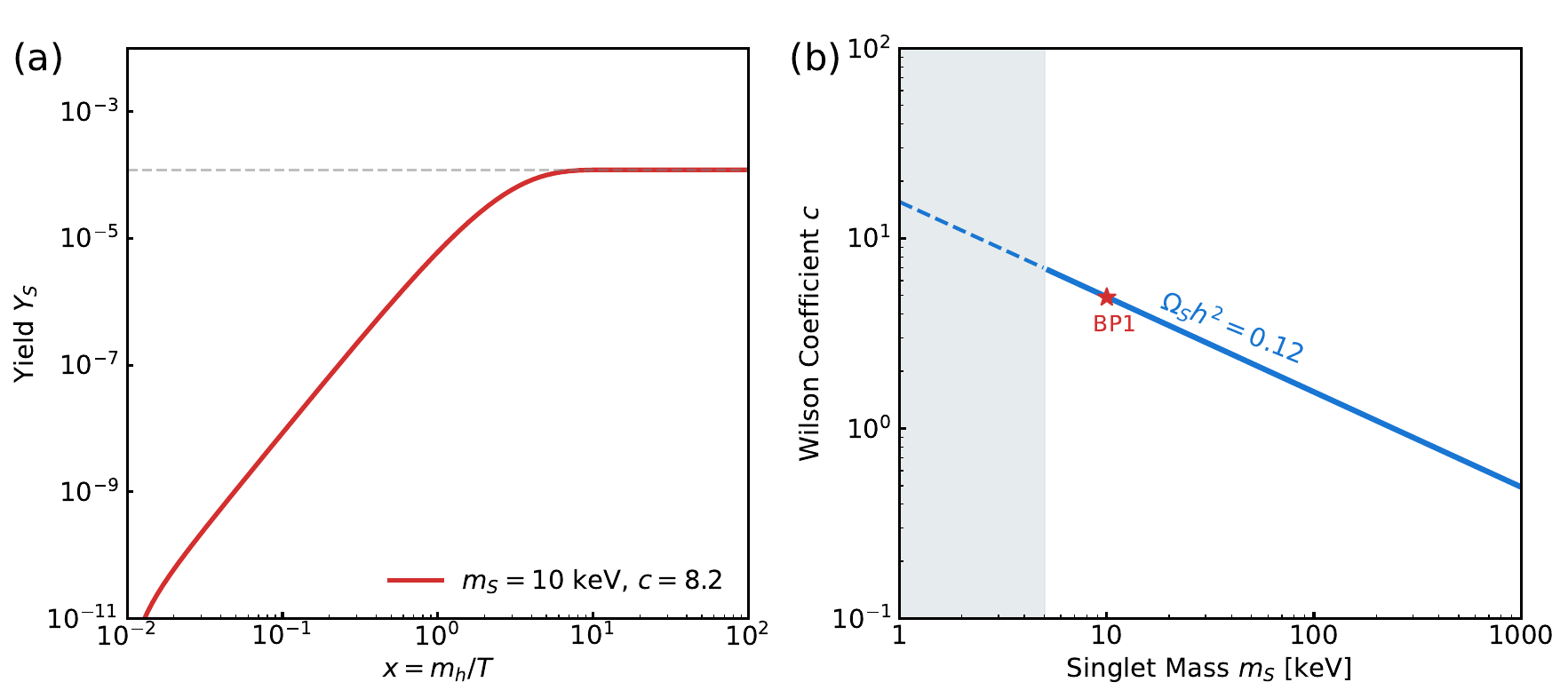}
    \caption{The yield evolution trajectory of the freeze-in dark matter and the global parameter space of the effective portal. The left panel illustrates the numerical evolution of the singlet yield $Y_S$ as a function of the inverse temperature parameter $x = m_h/T$, assuming benchmark values of $m_S = 10 \text{ keV}$ and $c = 5.0$. The right panel depicts the viable parameter space projected onto the singlet mass $m_S$ versus Wilson coefficient $c$ plane, where the solid blue line represents the theoretical solution that reproduces the observed dark matter relic abundance of $\Omega_S h^2 = 0.12$ \cite{cite1}. The transition to a dashed line indicates the boundary where the model enters the grey hatched region excluded by Lyman-$\alpha$ forest observations, with the red star designating the primary benchmark point.}
    \label{fig:freeze_in_results}
\end{figure}

The numerical evolution of the sterile singlet yield, illustrated in Figure \ref{fig:freeze_in_results}(a), demonstrates how the SM Higgs bosons in the thermal bath undergo rare decays that gradually populate the dark sector \cite{cite16}. The yield $Y_S$ originates from an initial value of zero and exhibits a characteristic S-shaped climb during the infrared freeze-in epoch \cite{cite5}. Once the temperature falls significantly below the Higgs mass, specifically at $x \approx 3$, the mother particle density is exponentially suppressed, leading to the cessation of the production process and the emergence of a constant asymptotic yield \cite{cite6, Bernal_2019}. This plateau reflects the final density of dark matter particles surviving in the expanding universe, confirming that the infrared regime is the dominant stage for relic generation in this framework.

To visualize the global viability of this mechanism, we project these numerical solutions onto the parameter space defined by the singlet mass $m_S$ and the Wilson coefficient $c$ of the dimension-six operator \cite{aebischer2022dark}. In Figure \ref{fig:freeze_in_results}(b), the blue contour signifies the trajectory where the total calculated abundance matches the precise relic density of $\Omega_S h^2 \approx 0.12$ reported by the Planck collaboration \cite{cite1}. It is evident that the observed dark matter density can be naturally achieved across a wide range of masses within the keV to MeV scale \cite{cite6}. Notably, our primary benchmark point BP1, situated at $m_S = 10$ keV, requires a Wilson coefficient of $c = 5.0$, which demonstrates that the observed abundance is a direct consequence of natural-order parameters within the effective field theory rather than extreme fine-tuning \cite{aebischer2022dark}.

Beyond the relic abundance, we must evaluate the impact of structural constraints on our warm dark matter candidate \cite{cite4}. Unlike standard CDM, a keV-scale singlet possesses a non-zero free-streaming length that effectively washes out small-scale density perturbations in the early universe \cite{cite4}. This suppression of the matter power spectrum is subject to rigorous testing via the Lyman-alpha forest spectra of distant quasars, which trace the distribution of neutral hydrogen in the intergalactic medium \cite{irvsivc2017new}. Observational data from high-resolution surveys place a robust lower bound on the mass of warm dark matter particles, as indicated by the shaded grey region in Figure \ref{fig:freeze_in_results}(b). While models falling below approximately 5 keV are astrophysically disfavored due to excessive smearing of cosmic structures, the benchmark point BP1 safely evades this exclusion zone \cite{irvsivc2017new}.

Radiative stability and early-universe consistency provide further layers of scrutiny for the sterile singlet. The scalar and lepton portals inevitably induce higher-order loop diagrams that mediate the radiative decay process $S \to \nu \gamma$, resulting in the emission of a monochromatic X-ray line with energy $E_\gamma = m_S / 2$ \cite{pal1982radiative}. The non-observation of such signatures by high-resolution telescopes, including XMM-Newton and Chandra, imposes a stringent upper bound on the effective mixing angle and the portal coupling constants \cite{roach2020nustar}. Similarly, the introduction of an additional light species must remain consistent with the successful predictions of Big Bang Nucleosynthesis (BBN) and the Cosmic Microwave Background (CMB) \cite{cite1}. Although the freeze-in mechanism prevents the singlet population from reaching full thermal equilibrium, its gradual production still adds a component to the effective number of relativistic degrees of freedom, $\Delta N_{eff}$ \cite{fields2020big}. Our calculations indicate that for the benchmark point $c = 5.0$, the resulting contribution to the radiation energy density is sufficiently suppressed, thereby maintaining the integrity of standard BBN predictions \cite{cite1, boyarsky2006constraints}.

The adoption of the ultra-high symmetry breaking scale $v_\phi \sim 10^{12}$ GeV is primarily motivated by the requirement to accommodate the QCD axion and resolve the strong CP problem. However, such a vast separation between the electroweak and the Peccei-Quinn scales introduces a potential hierarchy problem within the scalar sector, requiring careful stabilization of the Higgs mass \cite{gildener1976gauge, LIU2026102340}. To further explore the phenomenological versatility of our framework, we consider an alternative scenario where the symmetry breaking occurs at a lower scale, such as $v_\phi \sim 10$ TeV. In this low-energy regime, the $Z'$ gauge boson and exotic fermions would possess masses within the kinematic reach of current and near-future runs of the Large Hadron Collider (LHC). Specifically, latest dilepton resonance searches from the ATLAS and CMS collaborations have established stringent lower bounds on the $Z'$ mass, typically exceeding 4-5 TeV, which in turn constrains the available parameter space for the 10 TeV breaking scale \cite{aad2019search,alves2022constraining}.

Furthermore, the viability of the freeze-in mechanism at these lower energy scales is significantly modified by the altered interaction strengths. Since the effective portal coupling is inversely proportional to the suppression scale, a decrease in $v_\phi$ would necessitate a corresponding reduction in the fundamental Wilson coefficient $c$ to prevent the sterile singlet from reaching thermal equilibrium with the primordial plasma. If the interaction rates were to become sufficiently large, the dark matter would undergo standard freeze-out, which for a keV-scale singlet would lead to a severe overproduction of the relic density. Consequently, while the high-scale scenario provides a more natural explanation for the feebleness of the couplings via the large hierarchy, the low-scale variant offers the advantage of experimental falsifiability through direct collider searches for $Z'$ resonances and heavy scalar excitations. This complementarity ensures that the 3-3-1 axion framework remains a robust theoretical construction that can be scrutinized across a diverse range of cosmological and experimental frontiers.

\section{Conclusion}

We have demonstrated that embedding a sterile fermion singlet within the extended 3-3-1 electroweak framework yields a highly natural and multi-component dark matter paradigm. Driven by the exact local $Z_{13}$ discrete symmetry, the ultra-high Peccei-Quinn symmetry breaking scale dynamically suppresses the effective interaction of the singlet to the precise level required for non-thermal freeze-in production. This mechanism organically generates a keV-scale warm dark matter candidate capable of mitigating the persistent small-scale cosmological crises, such as the cusp-core and missing satellites anomalies. Concurrently, the very same symmetry breaking process produces the QCD axion, which naturally fulfills the role of the dominant CDM component necessary for seeding the large-scale cosmic structure. Ultimately, this unified framework elegantly shows that the complex dark matter landscape can emerge entirely from the fundamental algebraic protections of high-energy particle physics without relying on arbitrary parameter fine-tuning.

\appendix
\section{Complete Group Representations and Anomaly Cancellation}

To establish the theoretical robustness of the proposed dark matter framework, we must demonstrate that the introduction of the sterile singlet and the local discrete $Z_{13}$ symmetry does not spoil the anomaly-free nature of the underlying gauge structure. The standard 3-3-1 model successfully cancels all continuous gauge anomalies by balancing the number of $SU(3)_L$ triplets and anti-triplets across the fermion generations. In our extended model, the complete chiral fermion content and the scalar sector transform under the expanded group $\mathcal{G}_{3311} \otimes Z_{13}$. The specific representations and the parameterized discrete charge assignments are rigorously detailed in Table 1, where the continuous quantum numbers denote the transformations under $SU(3)_C$, $SU(3)_L$, $U(1)_X$, and $U(1)_N$, respectively.

\begin{table}[h!]
    \centering
    \renewcommand{\arraystretch}{1.4}
    \begin{tabular}{c c c}
        \hline\hline
        Field & $\mathcal{G}_{3311}$ Representation & $Z_{13}$ Charge \\
        \hline
        $Q_{aL} \; (a=1,2)$ & $(\mathbf{3}, \mathbf{3}^*, X_{Qa}, N_{Qa})$ & $q_{Qa}$ \\
        $Q_{3L}$ & $(\mathbf{3}, \mathbf{3}, X_{Q3}, N_{Q3})$ & $q_{Q3}$ \\
        $u_{iR}, d_{iR} \; (i=1,2,3)$ & $(\mathbf{3}, \mathbf{1}, X_{u,d}, N_{u,d})$ & $q_{ui}, q_{di}$ \\
        $J_{aR}, J_{3R}$ & $(\mathbf{3}, \mathbf{1}, X_{J}, N_{J})$ & $q_{Ja}, q_{J3}$ \\
        $L_{iL} \; (i=1,2,3)$ & $(\mathbf{1}, \mathbf{3}, X_{L}, N_{L})$ & $q_L$ \\
        $e_{iR} \; (i=1,2,3)$ & $(\mathbf{1}, \mathbf{1}, X_{e}, N_{e})$ & $q_{ei}$ \\
        $S$ & $(\mathbf{1}, \mathbf{1}, 0, 0)$ & $q_S$ \\
        $\eta, \rho, \chi$ & $(\mathbf{1}, \mathbf{3}, X_{\eta,\rho,\chi}, N_{\eta,\rho,\chi})$ & $q_\eta, q_\rho, q_\chi$ \\
        $\phi$ & $(\mathbf{1}, \mathbf{1}, 0, 0)$ & $q_\phi$ \\
        \hline\hline
    \end{tabular}
    \caption{Complete field content and anomaly-free charge assignments under the extended $\mathcal{G}_{3311} \otimes Z_{13}$ gauge symmetry.}
    \label{tab:representations}
\end{table}

For the local $Z_{13}$ discrete symmetry to be fundamentally consistent and immune to quantum violations, it must satisfy the Ibáñez-Ross anomaly cancellation conditions. These conditions mandate that the mixed triangular anomalies involving the discrete group and the continuous gauge or gravitational sectors must sum to zero modulo the discrete order $N=13$. The most stringent constraint arises from the mixed gravitational anomaly, denoted as $[\text{gravity}]^2 \times Z_{13}$, which requires the linear sum of all chiral fermion discrete charges to vanish. This condition is mathematically expressed as
\begin{equation}
    \mathcal{A}_{\text{grav}} = \sum_{\text{color}} \left( 2 q_{Qa} + q_{Q3} - \sum q_{ui} - \sum q_{di} - \sum q_{Ji} \right) + \sum_{\text{flavor}} \left( q_L - q_{ei} \right) + q_S \equiv 0 \pmod{13}.
\end{equation}
Similarly, the mixed strong interaction anomaly $[SU(3)_C]^2 \times Z_{13}$ specifically restricts the colored sector by evaluating the sum of the left-handed and right-handed quark charges multiplied by their respective Dynkin indices. This yields the subsequent algebraic constraint
\begin{equation}
    \mathcal{A}_{\text{color}} = 2 q_{Qa} + q_{Q3} - \sum_{i=1}^3 (q_{ui} + q_{di} + q_{Ji}) \equiv 0 \pmod{13}.
\end{equation}

The viability of our freeze-in dark matter scenario heavily relies on the coexistence of these global anomaly conditions with the specific local selection rules imposed on the scalar potential and the Yukawa sector. In the main text, we postulated that the generation of the Majorana mass for the singlet $S$ and its effective interaction via a dimension-six operator require the continuous gauge invariant terms to be strictly constrained by the discrete charges. Specifically, the fundamental Majorana coupling enforces the constraint
\begin{equation}
    2q_S + q_\phi \equiv 0 \pmod{13},
\end{equation}
while the effective dimension-six lepton portal $\frac{1}{\Lambda^2} (\overline{L_L} \eta) \phi^2 S$ rigidly selects the charge combination
\begin{equation}
    -q_L + q_\eta + q_S + 2q_\phi \equiv 0 \pmod{13}.
\end{equation}

By sequentially substituting the Majorana mass condition into the dimension-six operator requirement, we elegantly reduce the lepton portal constraint to $-q_L + q_\eta - 3q_S \equiv 0 \pmod{13}$. This crucial algebraic reduction demonstrates that the dark sector charge $q_S$ is fundamentally coupled to the visible lepton and scalar charges without generating any unresolvable mathematical contradictions. Furthermore, because the standard renormalizable Yukawa couplings of the 3-3-1 model generate a vast system of linear equations mapping the right-handed fermion charges to the scalar triplet charges, the colored sector contributions in the Ibáñez-Ross equations possess significant residual degrees of freedom. We can effortlessly absorb the constrained value of $q_S$ by appropriately shifting the unconstrained exotic quark charges $q_{Ji}$ within the modulo 13 arithmetic space. Consequently, the dimension-six operator prerequisite is perfectly self-consistent with all quantum anomaly cancellation conditions, definitively proving that the extreme weak coupling required for the FIMP miracle is a natural and robust mathematical consequence of the extended gauge topology.

\section{Scalar Potential Minimization and Exact Higgs-PQ Mixing}

To rigorously justify the effective mixing angle utilized in the cosmological production calculations, we present the explicit derivation of the scalar mass spectrum and the ensuing mixing dynamics. In the extended 3-3-1 framework incorporating the $Z_{13}$ discrete symmetry, the relevant scalar sector responsible for both electroweak and Peccei-Quinn symmetry breaking consists of the SM Higgs doublet $H$ and the complex scalar singlet $\phi$. The most general, renormalizable, and gauge-invariant scalar potential encompassing these two fields is formulated as
\begin{equation}
    V(H, \phi) = -\mu_H^2 H^\dagger H + \lambda_H (H^\dagger H)^2 - \mu_\phi^2 \phi^\dagger \phi + \lambda_\phi (\phi^\dagger \phi)^2 + \lambda_{h\phi} (H^\dagger H)(\phi^\dagger \phi),
\end{equation}
where $\mu_H^2$ and $\mu_\phi^2$ are the dimension-two mass parameters driving the spontaneous symmetry breaking, while $\lambda_H$, $\lambda_\phi$, and the portal coupling $\lambda_{h\phi}$ represent the dimensionless quartic couplings. 

Upon the spontaneous breaking of the respective symmetries, the scalar fields acquire their non-zero vacuum expectation values. We parameterize the fields around their classical minima in the unitary gauge as
\begin{equation}
    H = \begin{pmatrix} 0 \\ \frac{v_{EW} + h}{\sqrt{2}} \end{pmatrix}, \quad \phi = \frac{v_\phi + \rho}{\sqrt{2}} \exp\left(i \frac{a}{v_\phi}\right).
\end{equation}
In this parameterization, $v_{EW} \approx 246$ GeV characterizes the electroweak scale, $v_\phi \sim 10^{12}$ GeV represents the ultra-high Peccei-Quinn scale, $h$ and $\rho$ are the CP-even physical radial modes, and $a$ corresponds to the QCD axion. Substituting these expanded fields back into the scalar potential and demanding the first derivatives of the potential with respect to the vacuum expectation values to vanish, we obtain the fundamental minimization conditions, known as the tadpole equations. These conditions are explicitly given by
\begin{equation}
    \mu_H^2 = \lambda_H v_{EW}^2 + \frac{1}{2}\lambda_{h\phi} v_\phi^2, \quad \mu_\phi^2 = \lambda_\phi v_\phi^2 + \frac{1}{2}\lambda_{h\phi} v_{EW}^2.
\end{equation}

By utilizing these tadpole equations to eliminate the bare mass parameters $\mu_H^2$ and $\mu_\phi^2$, we can extract the quadratic terms governing the physical CP-even scalar masses. The cross-coupling term directly induces a mass mixing between the visible $h$ state and the dark $\rho$ state. The resulting $2 \times 2$ symmetric mass-squared matrix in the basis $(h, \rho)$ is derived as
\begin{equation}
    \mathcal{M}^2 = \begin{pmatrix} 2\lambda_H v_{EW}^2 & \lambda_{h\phi} v_{EW} v_\phi \\ \lambda_{h\phi} v_{EW} v_\phi & 2\lambda_\phi v_\phi^2 \end{pmatrix}.
\end{equation}
Because the off-diagonal elements are non-zero, the interaction eigenstates $(h, \rho)$ are not the true physical mass eigenstates. To find the physical propagating particles, we must diagonalize this matrix via an orthogonal rotation parameterized by a mixing angle $\theta$ \cite{cite22}.

The exact mixing angle required to diagonalize the mass matrix $\mathcal{M}^2$ is analytically determined by the standard relation
\begin{equation}
    \tan(2\theta) = \frac{2 \mathcal{M}^2_{12}}{\mathcal{M}^2_{22} - \mathcal{M}^2_{11}} = \frac{2 \lambda_{h\phi} v_{EW} v_\phi}{2\lambda_\phi v_\phi^2 - 2\lambda_H v_{EW}^2}.
\end{equation}
At this stage, we invoke the profound physical hierarchy inherent to our model, namely $v_{EW} \ll v_\phi$. Under this extreme scale separation, the diagonal entry corresponding to the heavy radial scalar mass vastly dominates the system, satisfying $\mathcal{M}^2_{22} \simeq 2\lambda_\phi v_\phi^2 \equiv m_\rho^2 \gg \mathcal{M}^2_{11}$. Consequently, we can safely perform a Taylor expansion in terms of the small ratio $v_{EW}/v_\phi$. In the small angle limit where $\tan(2\theta) \simeq 2\theta$, the mixing angle elegantly simplifies to the highly accurate approximation
\begin{equation}
    \theta \simeq \frac{\lambda_{h\phi} v_{EW} v_\phi}{m_\rho^2}.
\end{equation}
Correspondingly, the eigenvalues of the diagonalized matrix yield the physical masses of the observed 125 GeV Higgs boson and the heavy Peccei-Quinn scalar. Up to leading order in the perturbation expansion, the physical masses are naturally identified as $m_h^2 \simeq 2\lambda_H v_{EW}^2 - \theta^2 m_\rho^2 \approx 2\lambda_H v_{EW}^2$ and $m_\rho^2 \simeq 2\lambda_\phi v_\phi^2$. This rigorous algebraic derivation definitively confirms that the effective portal mixing angle utilized in our phenomenological freeze-in analysis is a robust and inescapable consequence of the underlying spontaneous symmetry breaking hierarchy.

\section{Phase Space Integration of the Boltzmann Equation}

To provide a rigorous foundation for the cosmological evolution of the sterile singlet, we detail the derivation of the yield equation from the first principles of kinetic theory. In the context of a Friedmann-Lemaître-Robertson-Walker universe, the evolution of the phase space distribution function $f_S(E, t)$ is governed by the Boltzmann equation, where the relativistic Liouville operator accounts for the expansion of the spacetime metric. The general form of this equation is expressed as
\begin{equation}
    \hat{L}[f_S] = E \frac{\partial f_S}{\partial t} - H |\vec{p}|^2 \frac{\partial f_S}{\partial E} = \mathcal{C}[f_S],
\end{equation}
where $H$ is the Hubble expansion rate and $\mathcal{C}[f_S]$ represents the collision integral encompassing all microscopic processes that create or annihilate the singlet particles. By integrating this expression over the momentum space and normalizing by the entropy density, we obtain the macroscopic evolution of the dark matter number density.

The production of the FIMP candidate in our model is dominated by the Higgs portal decay channel $h \to SS$. Because the effective coupling is extraordinarily small, the singlet population remains far from thermal equilibrium, allowing us to neglect the inverse annihilation process $SS \to h$ and simplify the occupancy distributions. Under the assumption that the plasma temperature is sufficiently low relative to the internal degrees of freedom, we adopt the Maxwell-Boltzmann approximation for the species in the thermal bath. In this limit, the distribution function for the Higgs boson simplifies significantly as
\begin{equation}
    f_h(E_h, T) \simeq \exp\left(-\frac{E_h}{T}\right),
\end{equation}
which provides a highly accurate description of the energy distribution during the critical freeze-in epoch near the electroweak scale.

The microscopic physics of the production process is encapsulated in the collision integral, which involves a multi-dimensional integration over the phase space of the participating species. Defining the Lorentz-invariant phase space element as $d\Pi_i = \frac{g_i d^3 p_i}{(2\pi)^3 2E_i}$, the rate of change of the singlet number density is formulated as
\begin{equation}
    \dot{n}_S + 3Hn_S = \int d\Pi_h d\Pi_{S_1} d\Pi_{S_2} (2\pi)^4 \delta^4(p_h - p_{S_1} - p_{S_2}) |\mathcal{M}|^2 f_h,
\end{equation}
where $|\mathcal{M}|^2$ is the squared transition matrix element summed over final states and averaged over initial states. By isolating the phase space of the decay products and utilizing the four-momentum delta function, the integral over $d\Pi_{S_1}$ and $d\Pi_{S_2}$ can be identified with the rest-frame decay width $\Gamma_{h \to SS}$ of the Higgs boson. Specifically, we utilize the kinematic identity
\begin{equation}
    \int d\Pi_{S_1} d\Pi_{S_2} (2\pi)^4 \delta^4(p_h - p_{S_1} - p_{S_2}) |\mathcal{M}|^2 = 2 m_h \Gamma_{h \to SS},
\end{equation}
which reduces the complex triple-species integral into a single integration over the momentum of the mother particle.

The remaining task involves performing the thermal averaging of the decay rate over the Higgs boson distribution in the expanding plasma. Substituting the identified decay width back into the Boltzmann equation and switching the integration variable from momentum to energy $E_h$, the production rate density is expressed as
\begin{equation}
    \dot{n}_S + 3Hn_S = \frac{g_h m_h \Gamma_{h \to SS}}{2\pi^2} \int_{m_h}^\infty \sqrt{E_h^2 - m_h^2} \exp\left(-\frac{E_h}{T}\right) dE_h.
\end{equation}
By performing a change of variables such that $E_h = m_h \cosh(\theta)$, this integral is analytically recognized as the definition of the modified Bessel function of the second kind of order one. Specifically, we arrive at the simplified macro-rate
\begin{equation}
    \dot{n}_S + 3Hn_S = \frac{g_h m_h^2 \Gamma_{h \to SS}}{2\pi^2} T K_1\left(\frac{m_h}{T}\right).
\end{equation}
Finally, by expressing the time derivative in terms of the dimensionless inverse temperature $x = m_h/T$ and normalizing by the entropy density $s$, we obtain the definitive yield equation $dY_S/dx$ presented in the main text. This systematic reduction proves that the macroscopic freeze-in dynamics are rigorously coupled to the microphysical decay properties through the $K_1(x)$ kernel, ensuring a self-consistent treatment of the non-thermal dark matter genesis.

\acknowledgments

This work is supported by the Developing Project of Science and Technology of Jilin Province (20250102032JC). 

\section*{Data Availability Statement}
The data underlying this article will be shared at request to the authors.




\begin{thebibliography}{10}

\bibitem{cite1}
N.~Aghanim et~al., \emph{Planck 2018 results. vi. cosmological parameters}, {\emph{Astron. Astrophys} {\bfseries 641} (2020) A6}.

\bibitem{cite2}
J.S.~Bullock and M.~Boylan-Kolchin, \emph{Small-scale challenges to the $\lambda$ cdm paradigm}, {\emph{Annual Review of Astronomy and Astrophysics} {\bfseries 55} (2017) 343}.

\bibitem{cite4}
M.~Viel, J.~Lesgourgues, M.G.~Haehnelt, S.~Matarrese and A.~Riotto, \emph{Constraining warm dark matter candidates including sterile neutrinos and light gravitinos with wmap and the lyman-$\alpha$ forest}, {\emph{Physical Review D—Particles, Fields, Gravitation, and Cosmology} {\bfseries 71} (2005) 063534}.

\bibitem{cite5}
L.J.~Hall, K.~Jedamzik, J.~March-Russell and S.M.~West, \emph{Freeze-in production of fimp dark matter}, {\emph{Journal of High Energy Physics} {\bfseries 2010} (2010) 1}.

\bibitem{cite6}
N.~Bernal, M.~Heikinheimo, T.~Tenkanen, K.~Tuominen and V.~Vaskonen, \emph{The dawn of fimp dark matter: a review of models and constraints}, {\emph{International Journal of Modern Physics A} {\bfseries 32} (2017) 1730023}.

\bibitem{cite9}
P.H.~Frampton, \emph{Chiral dilepton model and the flavor question}, {\emph{Physical Review Letters} {\bfseries 69} (1992) 2889}.

\bibitem{LIU2026102340}
F.~Liu and F.~Liu, \emph{Gravitational robustness of automatic peccei-quinn symmetry in 3-3-1 models}, \href{https://doi.org/https://doi.org/10.1016/j.dark.2026.102340}{\emph{Physics of the Dark Universe} (2026) 102340}.

\bibitem{cite11}
P.V.~Dong, T.D.~Tham and H.T.~Hung, \emph{3-3-1-1 model for dark matter}, \href{https://doi.org/10.1103/PhysRevD.87.115003}{\emph{Phys. Rev. D} {\bfseries 87} (2013) 115003}.

\bibitem{cite13}
L.E.~Ibáñez and G.G.~Ross, \emph{Discrete gauge symmetries and the origin of baryon and lepton number conservation in supersymmetric versions of the standard model}, \href{https://doi.org/https://doi.org/10.1016/0550-3213(92)90195-H}{\emph{Nuclear Physics B} {\bfseries 368} (1992) 3}.

\bibitem{cite14}
G.G.~Raffelt, \emph{Astrophysical axion bounds},  in \emph{Axions: Theory, Cosmology, and Experimental Searches}, pp.~51--71, Springer (2008).

\bibitem{cite15}
P.~Bode, J.P.~Ostriker and N.~Turok, \emph{Halo formation in warm dark matter models}, {\emph{The Astrophysical Journal} {\bfseries 556} (2001) 93}.

\bibitem{Bernal_2019}
N.~Bernal, F.~Elahi, C.~Maldonado and J.~Unwin, \emph{Ultraviolet freeze-in and non-standard cosmologies}, \href{https://doi.org/10.1088/1475-7516/2019/11/026}{\emph{Journal of Cosmology and Astroparticle Physics} {\bfseries 2019} (2019) 026}.

\bibitem{aebischer2022dark}
J.~Aebischer, W.~Altmannshofer, E.E.~Jenkins and A.V.~Manohar, \emph{Dark matter effective field theory and an application to vector dark matter}, {\emph{Journal of High Energy Physics} {\bfseries 2022} (2022) 86}.

\bibitem{ibanez1992discrete}
L.E.~Ibanez and G.G.~Ross, \emph{Discrete gauge symmetries and the origin of baryon and lepton number conservation in supersymmetric versions of the standard model}, {\emph{Nuclear Physics B} {\bfseries 368} (1992) 3}.

\bibitem{cite16}
B.~Patt and F.~Wilczek, \emph{Higgs-field portal into hidden sectors}, {\emph{arXiv preprint hep-ph/0605188} (2006) }.

\bibitem{cite17}
O.~Lebedev, H.M.~Lee and Y.~Mambrini, \emph{Vector higgs portal dark matter and the invisible higgs}, {\emph{Physics Letters B} {\bfseries 707} (2012) 570}.

\bibitem{cite18}
M.~Frigerio, T.~Hambye and E.~Masso, \emph{Sub-gev dark matter as pseudo-nambu-goldstone bosons from the seesaw scale}, {\emph{Physical Review X} {\bfseries 1} (2011) 021026}.

\bibitem{cite20}
K.~Saikawa and S.~Shirai, \emph{Primordial gravitational waves, precisely: The role of thermodynamics in the standard model}, {\emph{Journal of Cosmology and Astroparticle Physics} {\bfseries 2018} (2018) 035}.

\bibitem{cite22}
P.~Gondolo and G.~Gelmini, \emph{Cosmic abundances of stable particles: Improved analysis}, {\emph{Nuclear Physics B} {\bfseries 360} (1991) 145}.

\bibitem{irvsivc2017new}
V.~Ir{\v{s}}i{\v{c}}, M.~Viel, M.G.~Haehnelt, J.S.~Bolton, S.~Cristiani, G.D.~Becker et~al., \emph{New constraints on the free-streaming of warm dark matter from intermediate and small scale lyman-$\alpha$ forest data}, {\emph{Physical Review D} {\bfseries 96} (2017) 023522}.

\bibitem{pal1982radiative}
P.B.~Pal and L.~Wolfenstein, \emph{Radiative decays of massive neutrinos}, {\emph{Physical Review D} {\bfseries 25} (1982) 766}.

\bibitem{roach2020nustar}
B.M.~Roach, K.C.~Ng, K.~Perez, J.F.~Beacom, S.~Horiuchi, R.~Krivonos et~al., \emph{Nustar tests of sterile-neutrino dark matter: new galactic bulge observations and combined impact}, {\emph{Physical Review D} {\bfseries 101} (2020) 103011}.

\bibitem{boyarsky2006constraints}
A.~Boyarsky, A.~Neronov, O.~Ruchayskiy and M.~Shaposhnikov, \emph{Constraints on sterile neutrinos as dark matter candidates from the diffuse x-ray background}, {\emph{Monthly Notices of the Royal Astronomical Society} {\bfseries 370} (2006) 213}.

\bibitem{fields2020big}
B.D.~Fields, K.A.~Olive, T.-H.~Yeh and C.~Young, \emph{Big-bang nucleosynthesis after planck}, {\emph{Journal of Cosmology and Astroparticle Physics} {\bfseries 2020} (2020) 010}.

\bibitem{gildener1976gauge}
E.~Gildener, \emph{Gauge-symmetry hierarchies}, {\emph{Physical Review D} {\bfseries 14} (1976) 1667}.

\bibitem{aad2019search}
G.~Aad, B.~Abbott, D.C.~Abbott, A.A.~Abud, K.~Abeling, D.~Abhayasinghe et~al., \emph{Search for high-mass dilepton resonances using 139 fb- 1 of pp collision data collected at s= 13 tev with the atlas detector}, {\emph{Physics Letters B} {\bfseries 796} (2019) 68}.

\bibitem{alves2022constraining}
A.~Alves, L.~Duarte, S.~Kovalenko, Y.~Oviedo-Torres, F.~Queiroz and Y.~Villamizar, \emph{Constraining 3-3-1 models at the lhc and future hadron colliders}, {\emph{Physical Review D} {\bfseries 106} (2022) 055027}.

\end{thebibliography}


\providecommand{\href}[2]{#2}\begingroup\raggedright\endgroup

\end{document}